# Data Descriptor Scope

## Title
*DATA REPOSITORY OF FINITE ELEMENT MODELS OF NORMAL AND DEFORMED THORACOLUMBAR SPINE*

## Authors
Morteza Rasouligandomani[1], Alex del Arco[2], Francis Kiptengwer Chemorion [1,3,4], Marc-Antonio Bisotti [3], Fabio Galbusera[5], Jérôme Noailly[1], Miguel A. González Ballester[1,6]

## Affiliations
1. BCN MedTech, Department of Information and Communication Technologies, Universitat Pompeu Fabra, Barcelona, Spain
2. Hospital del Mar, Barcelona, Spain
3. InSilicoTrials Technologies, Trieste, Italy
4. Barcelona Supercomputing Center, Barcelona, Spain
5. IRCCS Galeazzi, Milan, Italy; Schulthess Klinik, Zürich, Switzerland
6. ICREA, Barcelona, Spain
Corresponding authors: Morteza Rasouligandomani (morteza.rasouli@upf.edu), and Jérôme Noailly (jerome.noailly@upf.edu)

## Abstract
Adult spine deformity (ASD) is prevalent and leads to a sagittal misalignment in the vertebral column. Computational methods, including Finite Element (FE) Models, have emerged as valuable tools for investigating the causes and treatment of ASD through biomechanical simulations. However, the process of generating personalized FE models is often complex and time-consuming. To address this challenge, we present a repository of FE models with diverse spine morphologies that statistically represent real geometries from a cohort of patients. These models are generated using EOS images, which are utilized to reconstruct 3D surface spine models. Subsequently, a Statistical Shape Model (SSM) is constructed, enabling the adaptation of a FE hexahedral mesh template for both the bone and soft tissues of the spine through mesh morphing. The SSM deformation fields facilitate the personalization of the mean hexahedral FE model based on sagittal balance measurements. Ultimately, this new hexahedral SSM tool offers a means to generate a virtual cohort of 16807 thoracolumbar FE spine models, which are openly shared in a public repository.

## Background & Summary

The Finite Element (FE) Method is well-known for the numerical solving of differential equations through the discretization of complex geometrical domains into discrete cells, or elements, in which physics-based partial differential equations can solved. FE analyses have shown their value to explore causatively spine pathologies, or spine surgical treatments and devices [1,2,3,4]. Recently, several studies focused on FE simulations of lumbar spine deformity (only scoliosis), spine surgery, and mechanical response of Intervertebral Disk (IVD) and soft tissues. However, there was lack of biomechanical FE models for thoracolumbar spine sagittal deformities.

As spine sagittal deformity is a patient-specific deviation from the standard spine geometry, the modelling thereof needs to be personalised. Clinical images of the lumbar and thoracolumbar spine have been exploited in patient-specific FE modelling of the spine [5,6,7,8]. Furthermore, image analysis techniques such as statistical shape analysis have led to Statistical Shape Models (SSM), providing a powerful tool to describe morphological variations in a specific population and use this knowledge to personalize the models according to specific patient characteristics. SSM has been proposed for the lumbar, thoracic, and cervical spines [9,10,11,12,13]. Yet, spine sagittal deformity and morphological balance explorations require the modelling of, at least, the thoracolumbar spine including the pelvis and sacrum, and the femoral head [14].

In biomechanical FE models, such as spine FE model, shapes need to be discretized, or meshed, into elements for the iterative solving of the partial differential equations that mathematically describe the physics of the mechanically loaded organ. The meshes can be classified in two main categories: triangulated and hexahedral. In contrast to triangulated meshes, hexahedral meshes follow the structural organization of the tissues that build the organ. They are particularly convenient to define the local anisotropic material properties or oriented material reinforcements, found in biological tissues [15].

Hexahedral meshes are the most accurate FE meshes which allow optimal mesh convergence for the numerical solving of systems of partial differential equations. This becomes particularly evident under large deformations, e.g., as experienced by spine intervertebral discs [16-17]. However, their automatic adaptation to nonlinear tissue and organ shapes is challenging and require cumbersome manual operations, especially when used in structural meshes. As automatic segmentations of medical images provide triangulated surface meshes, the creation of 3D structural meshes out of these segmentations stands for a strong limitation, therefore, to define numerically optimal patient-specific spine FE models. Recently, a few studies focused on the thoracolumbar spine FE hexahedral meshes, in which no sagittal deformity had been observed.

We propose to overcome this limitation, as we report an automatable pipeline to build hexahedral thoracolumbar FE spine models out of triangulated surface meshes. Hereby, the triangulated meshes were obtained from EOS 3D information, to allow the modelling of an extensive anatomical region, including the thoracolumbar spine, the sacrum, the pelvis, and the femoral head. We used SSM and the morphing of pre-existing structural meshes developed and verified for the osteo-ligamentous spine [18], to eventually achieve personalized models that can be used to explore the biomechanics of the thoracolumbar spine, both in balanced and adult spine sagittal deformity. Thanks to the SSM, the full FE modelling pipeline (Figure 1) allowed to create a virtual cohort of 16807 models through the coupling of different shape modes.

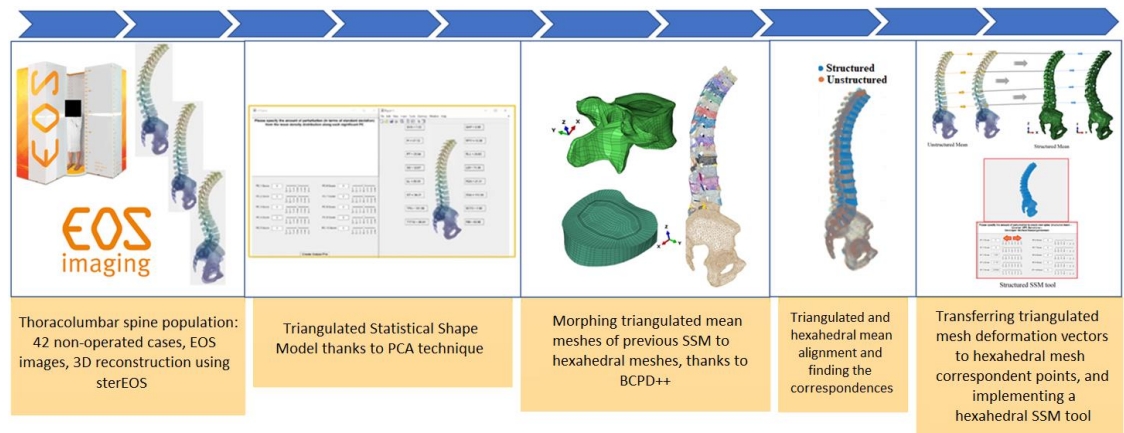

Figure 1: Pipeline to generate FE thoracolumbar spine meshes

As such, the aim of this article is to share an extensive repository of ready-to-use structural FE models of the thoracolumbar spine, including (i) a set models personalised against real medical images, and (ii) a set of models that stand for a virtual FE cohort. The model database is further annotated according to state-of-the-art clinical parameters used to quantify the spine geometry and sagittal balance. Overall, the bi-planar EOS images of 42 pre-operated patients affected by adult spine deformity were gathered and converted into 3D surface shape models aligned by rigid registration. Then, Principal Component Analyses (PCA) was used to reduce the dimensionality and explore shape modes representing shape variabilities, as well as computing a mean model. The latter was subsequently transformed to a hexahedral FE model with a 3D hexahedral mesh, using mesh morphing techniques ([19], [20]). At this stage, the IVD and ligament meshes were included, and the SSM deformation fields that resulted from the PCA were applied on the mean osteo-ligamentous spine hexahedral mesh, to achieve the shared virtual spine deformity thoracolumbar FE models, plus a set of well-balanced spine models.

There are several existing repositories for spine FE models. SpineWeb is a repository which shares lumbar and cervical FE models, developed by the University of California, Berkeley [21]. Vertebral Model Repository (VMR) is another repository which distributes computational models of individual vertebrae, developed by the University of Auckland, New Zealand. Additionally, another repository shares two patient-specific (49y.o., and 59y.o., Female) structural FE meshes of the lumbar spine (L1-L5) and models of functional spine units (FSUs) [22]. These FE models were developed out of segmented CT image data, including IVDs and ligaments. Furthermore, lumbosacral spine models were shared in ArtiSynth [23]. This repository holds hexahedral calibrated and validated models of lumbosacral spine including IVDs, ligaments and sacrum, built with the freely available 3D modelling platform ArtiSynth which supports the combined simulation of multibody and FE models. The existing repositories mentioned earlier suffer from certain limitations. They offer only a limited number of patient-specific models, lack validated thoracolumbar osteo-ligamentous FE models, and do not provide sagittal deformity quantification. As a result, there is a need for a comprehensive repository that addresses these shortcomings. Such a repository would offer a substantial collection of thoracolumbar osteo-ligamentous FE models, incorporating sagittal deformity quantification, thereby providing a valuable resource in this field.

To the best of our knowledge, the thoracolumbar spine FE model repository presented in this article is unique. It comprises 16807 3D thoracolumbar spine FE models, generated through statistical shape variations, representing a virtual cohort. These models have been diligently annotated in terms of spine sagittal balance and are made available through the Zenodo repository. Notably, the quality of the mesh and the ability of the FE models to accurately reproduce experimentally measured Ranges of Motion (ROM) were thoroughly evaluated and reported.

## Methods
### A. Thoracolumbar spine triangulated statistical shape model
Here, we used Shape Statistical Modelling (SSM) [24] to analyse and capture the variations within a shape family, as well as generate new shapes.

#### Thoracolumbar spine population
In this study, 42 non-operated EOS bi-planar images are included from IRCCS (Orthopaedic Hospital of Galeazzi, Milan). Inclusion criteria are:
- Age: 50 to 75 y.o.
- Disease: Spine sagittal deformity
- (LL – PI) > 10º
- PT > 20º
- SVA > 5 cm

A total of 42 patient-personalized models of the thoracolumbar spine were reconstructed using the sterEOS software. The sterEOS software utilizes the EOS™ X-ray machine from EOS imaging company in Paris, which has the capability to simultaneously capture bi-planar X-ray images. This innovative technology allows for scanning the entire body in an upright, load-bearing position while minimizing radiation exposure with ultra-low doses [25]. The reconstruction process of the 3D spine surfaces involves two steps, as described below:

1. **Identification of Control Points:** A specialist user selects eight control points located at the centre of the vertebral bodies in both sagittal and frontal X-ray images. These control points serve as reference points for estimating and drawing the contours of the vertebrae in the two planes. This initial step results in the reconstruction of a preliminary 3D thoracolumbar spine model using triangulated meshes (Figure 2, step 1).
2. **Adjustment of Vertebrae Contours:** The user can modify the position and orientation of the vertebrae contours to ensure they align accurately with the corresponding bi-planar X-ray images. This step allows for fine-tuning the reconstructed model to achieve a more precise representation of the patient's spine (Figure 2, step 2).

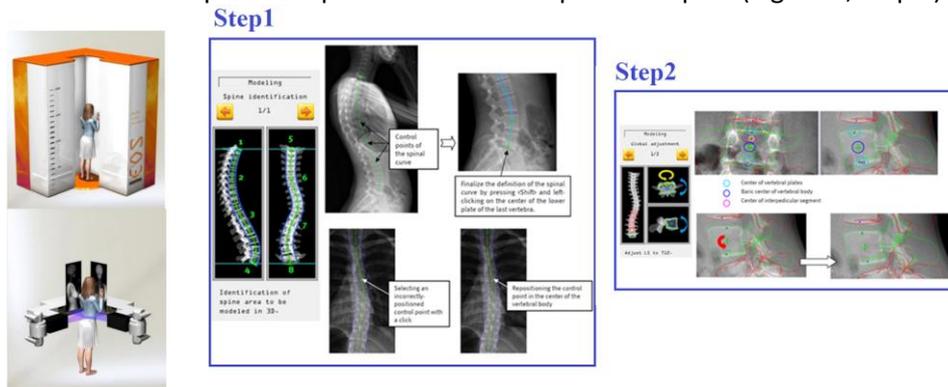

Figure 2: 3D reconstruction of thoracolumbar spine models (triangulated meshes) using sterEOS

42 thoracolumbar spine surface meshes have been saved in the standard DICOM (Digital Imaging and Communications in Medicine) format. To facilitate further analysis and processing, all the 3D DICOM files were converted to the stl format using a C++ code compiled in the Fedora Linux environment. The conversion process was conducted using QT5 Creator and the Mingw environment, with the DICOM to stl converter provided by IRCCS, Milan. Consequently, 42 stl files representing patient-personalized models of the thoracolumbar spine based on EOS imaging were generated. These stl files are available in a public repository. Additionally, these stl files were utilized to train the Shape Space Model (SSM). The spinopelvic parameters for 42 patient-personalized stl files were measured and documented in the accompanying Excel file. 42 triangulated models were reconstructed in their respective image frames; hence, it is necessary to align all the shapes for further analysis and comparison.

*Generalized Procrustes Alignment (GPA)*

All of 42 shapes are aligned rigidly using Generalized Procrustes Alignment (GPA) [26]. Following steps are used for GPA:
- Choosing an arbitrary shape as a reference shape
- Aligning all the shapes to the reference shape
- Computing the mean shape for the aligned shapes
- Considering the mean shape as the new reference shape, and iterating

After aligning the training sets, it is needed to find the correspondences between the points in different models, but all the points in the 42 models are correspondent to each other thanks to EOS imaging system.

*Developing thoracolumbar spine triangulated SSM*

Principal Component Analysis (PCA) is a widely used technique in shape modelling that is employed in this study to reduce dimensionality and obtain a compact representation of shape variability [27]. PCA serves as a fundamental tool by extracting the principal modes of variation through the computation of eigenvectors derived from the covariance matrix. The covariance matrix, which encapsulates the amount of variation present in the dataset relative to the mean shape, is defined by Equation 1 [27].

$$Cov = \frac{1}{n-1}\sum_{i=1}^{n}(X_i - \mu)(X_i - \mu)^T \quad (1)$$

where n is the number of samples, $X_i$ is a matrix including X, Y and Z point coordinates in each sample, and $\mu$ stands for the average shape which is introduced in Equation 2.

$$Mean = \frac{1}{n}\sum_{i=1}^{n} X_i \quad (2)$$

Eigenvectors of the covariance matrix are obtained using Singular Value Decomposition (SVD), an operation to decompose a matrix into three matrices, two unitary (eigenvectors) and one diagonal (eigenvalue) matrices. So, the covariance matrix can be decomposed into three matrices in Equation 3.

$$X = \emptyset S \emptyset^T \quad (3)$$

where $\emptyset$ is matrix of eigenvectors, and S is the diagonal matrix of eigenvalues. Then, shapes can be expressed in Equation 4 as a linear combination of eigenvectors given the feature weights $w_i$ [27].

$$\text{Shape} = \text{Mean} + \sum_{i=1}^{m} \sqrt{d_i}\,\emptyset_i w_i \quad (4)$$

where m is the number of activated shape modes, $d_i$ is $i^{th}$ diagonal element of matrix S in Equation 3, $\emptyset_i$ is $i^{th}$ column of matrix $\emptyset$ in Equation 3, and $w_i$ is the standard deviation (SD) weight coefficient of activated shape mode, typically $w_i \epsilon \{-3,3\}$. Hence, each mode in the SSM represents a specific direction of shape variation, which can be visualized as movements of landmarks along straight lines defined by the eigenvectors passing through their mean position. In this study, the eigenvectors and eigenvalues of the covariance matrix are computed using the PCA technique for the set of 42 aligned and corresponding non-operated thoracolumbar spine models. The SSM is then constructed using Equation 4.

Furthermore, we used compactness analysis for determining enough shape modes. Compactness analysis involves assessing the cumulative variance (Equation 5) for different numbers of modes. To determine the optimum number of activated shape modes (m, as defined in Equation 4), a threshold, such as 95%, is commonly used. The threshold indicates the desired level of explained variance in the training dataset.

$$\text{Cumulative variance} = \frac{\sum_{j=1}^{m} d_j}{\textit{summation of all eigenvalues}} \quad (5)$$

At the end, 16807 thoracolumbar spine triangulated models (stl files) were sampled using the triangulated SSM tool. The sampling process involved combining the first 5 Principal Components (PCs) and adjusting seven standard deviations (SDs) (-3, -2, -1, 0, 1, 2, 3) for each shape mode. The sampled data, along with the corresponding spinopelvic measurements, are stored in a publicly accessible repository for further analysis and research.

## B. Thoracolumbar osteo-ligamentous spine FE hexahedral statistical shape model

Following the pipeline depicted in Figure 1, the mean shape of the SSM in triangulated meshes is transformed into hexahedral meshes using mesh morphing techniques. To achieve the morphed mean model, the morphing process was applied to both vertebrae and intervertebral discs (IVDs). The ligaments were incorporated into the model, resulting in the generation of the morphed mean model. Later, the thoracolumbar osteo-ligamentous spine hexahedral SSM was presented.

### Vertebra mesh morphing

The long instrumentation in spine surgery typically involves a minimum of four levels of distal fusion, extending from S1 to L3 or T1. Vertebra fracture and screw loosening at the upper instrumented vertebra are important biomechanical parameters [28]. To address this, the anterior parts of vertebras L3 to T1 in the mean triangulated meshes were converted to hexahedral meshes using the accelerated Bayesian Coherent Point Drift (BCPD++) method. The posterior parts of vertebras L3 to L1 and L5 and L4 meshes were converted to quadrilateral meshes using BCPD++. The posterior parts of vertebras T12 to T1 were kept as triangulated elements. Consequently, two template meshes were required: L3 hexahedral meshes for the anterior part (to morph anterior L3 to T1) and L3 quadrilateral meshes (to morph posterior L3 to L1, and entire L5 and L4). The hexahedral meshes were also extruded in the pedicle vertebra using Abaqus Simulia, depending on the pedicle diameter.

BCPD++ is an extension of the Bayesian Coherent Point Drift (BCPD) algorithm [29, 30]. BCPD is a probabilistic method for point set registration [31], as an extension of the Coherent Point Drift (CPD) algorithm [20]. The CPD algorithm models the transformation between two-point sets using a Gaussian Mixture Model (GMM) and iteratively solves for the optimal transformation parameters [20]. The GMM consists of K Gaussian distributions in the transformation space, and the probability density function of the GMM is defined as [31]:

$$P(Y|X, \Theta) = \sum_{j=1}^{K} w_j \times N(Y|\mu_j, \Sigma j) \quad (6)$$

where Y is the transformed point set, X is the reference point set, $\Theta = \{w_j, \mu_j, \Sigma j\}$ is the set of parameters of the GMM, and $N(\mu_j, \Sigma j)$ is the Gaussian distribution with mean $\mu_j$ and covariance matrix $\Sigma j$. Then, BCPD computes the correspondence probability between a point x in the reference point set and a point y in the transformed point set as [31]:

$$P(x, y|X, Y, \Theta) = w_j \times N(y|\mu_j, \Sigma j) \quad (7)$$

where j is the index of the nearest centroid to y; and $w_j$, $\mu_j$, and $\Sigma j$ are the parameters of the corresponding Gaussian distribution. Then, BCPD updates the GMM by re-estimating the centroid locations and the covariances of the Gaussian distributions. It also estimates the posterior distribution of the transformation parameters using Markov Chain Monte Carlo (MCMC) methods. The posterior distribution is given by Bayes' rule as:

$$P(\Theta|X, Y) = P(Y|X, \Theta)P(\Theta)/P(Y|X) \quad (8)$$

where $P(\Theta)$ is the prior distribution of the parameters, and $P(Y|X)$ is the normalization constant. The posterior distribution in BCPD provides valuable information about the uncertainty associated with the estimated transformation parameters. By incorporating a prior distribution, usually a Gaussian distribution, on the transformation parameters, the algorithm becomes more robust against noise and outliers. The estimated posterior distribution is then utilized in BCPD to further refine the transformation parameters, either by computing the mean or the mode of the distribution. The BCPD algorithm iterates until convergence is achieved, or a maximum number of iterations is reached. This iterative process allows for the optimization of the transformation parameters based on the available data and the underlying statistical models.

In this study, the BCPD++ open-source codes are downloaded from the GitHub repository[32] and used to perform the morphing of the template vertebra. The morphing process utilizes a multi-layer BCPD++ approach, where if the initial morphing yields a Euclidean distance greater than 0.04 mm, the prior output is considered as the template mesh, and the BCPD++ parameters are fine-tuned again. This process is repeated until the Euclidean distance falls below 0.04 mm. To enhance performance and reduce computational time, an initial rotation and rigid registration are applied to the source. The template mesh undergoes a 90º rotation about the y-axis for the initial rotation. Rigid registration is performed using the available BCPD++ code[32] with a sufficiently large lambda value, such as 1e9. Lambda is one of the tuned parameters in BCPD++, controlling the expected length of deformation vectors. For the BCPD++ non-rigid registration, the initial parameters are specified as shown in Table 1.

Table 1: Initial parameters for BCPD++ non-rigid registration

| Parameters | Values |
|---|---|
| Interaction between the points, Lambda | 2 |
| Motion smoothing weight, Lambda | 8 |
| Randomness of the point matching at the beginning of the optimization, g | 2 |
| Nystrom samples for computing G, K | 70 |
| Nystrom samples for computing P, J | 300 |
| Scale factor of sigma that defines areas to search for neighbours, d | 7 |
| Maximum radius to search for neighbours, e | 0.2 |
| The value of sigma at which the KD tree search is turned on, f | 0.2 |
| Downsample radius, r | 0.5 |
| Expected percentage of outliers, Outlier Ratio | 0.1 |
| Maximum number of iterations | 1000 |
| Tolerance between consecutive CPD iterations | 1e-15 |

Therefore, the morphing process involves converting 15 anterior triangulated vertebras (from L3 to T1) into hexahedral meshes using BCPD++, while 3 posterior triangulated vertebras (L3 to L1) and 2 entire L4 and L5 triangulated vertebras are morphed into quadrilateral meshes using BCPD++. Additionally, hexahedral meshes are extruded in the pedicle region. The sacrum or pelvis, as well as the posterior elements of T1 to T12, remain as triangulated meshes. The vertebra mesh morphing process is presented in Figure 3.

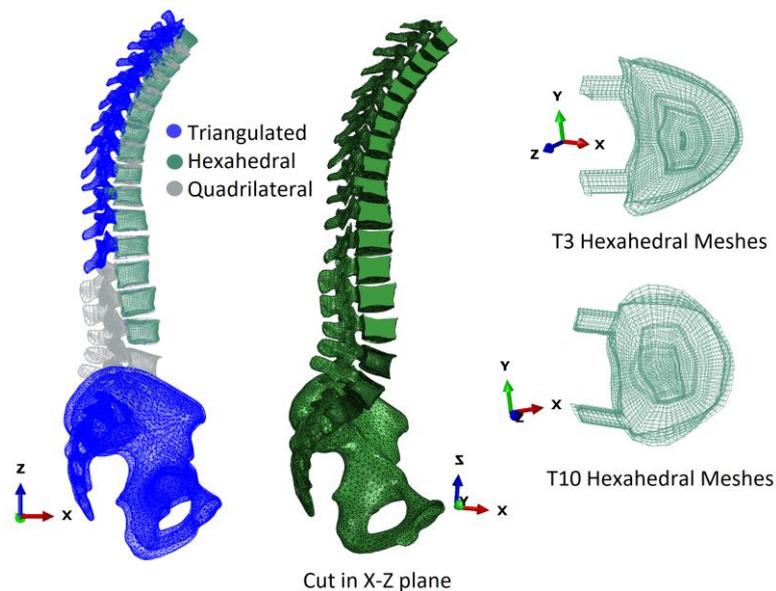

Figure 3: vertebra mesh morphing process

*IVD mesh morphing*

The IVD surface meshes are not included in the stl files; therefore, quadrilateral meshes are projected onto the two adjacent endplates of the mean morphed vertebras. The outer endplate's adjacent points are connected to create surrounded quadrilateral elements. These elements are then divided into 8-layer quadrilateral meshes. This process generates the outer IVD quadrilateral meshes in Abaqus software (Simulia). The mean outer IVD quadrilateral meshes are subsequently converted to hexahedral meshes using the BCPD++ method.

For the IVD mesh morphing, an IVD template hexahedral mesh (L4/L3) is selected and morphed to the target IVD structure (shown in Figure 4). The IVD template mesh includes the Annulus Fibrosus (AF), Nucleus Pulposus (NP), and Cartilage Endplate (CEP). The average thickness of the CEP in the IVD template is 0.7mm, and the NP proportion is 40%. The BCPD++ method is used to morph the template IVD. Initially, the AF and NP are morphed individually and then integrated to create a volumetric fine mesh meta model. Subsequently, the IVD hexahedral template, including the CEP, is morphed to match the volumetric meta model while ensuring that the CEP thickness remains within an acceptable range (0.5 to 1.5 mm). The multi-layer BCPD++ mesh refinement technique, like the vertebra morphing process, is utilized for IVD morphing. Similarly, an initial rotation of 90º about the y-axis is applied to the IVD template mesh. BCPD++ rigid registration with a large lambda value (e.g., 1e9) is used to roughly align the IVD target and template. BCPD++ non-rigid registration is then performed using the BCPD++ initial parameters from Table 1.

Consequently, 17 IVD hexahedral meshes are morphed and integrated with the vertebras. The CEP thickness and NP proportions are controlled to ensure they fall within an acceptable range.

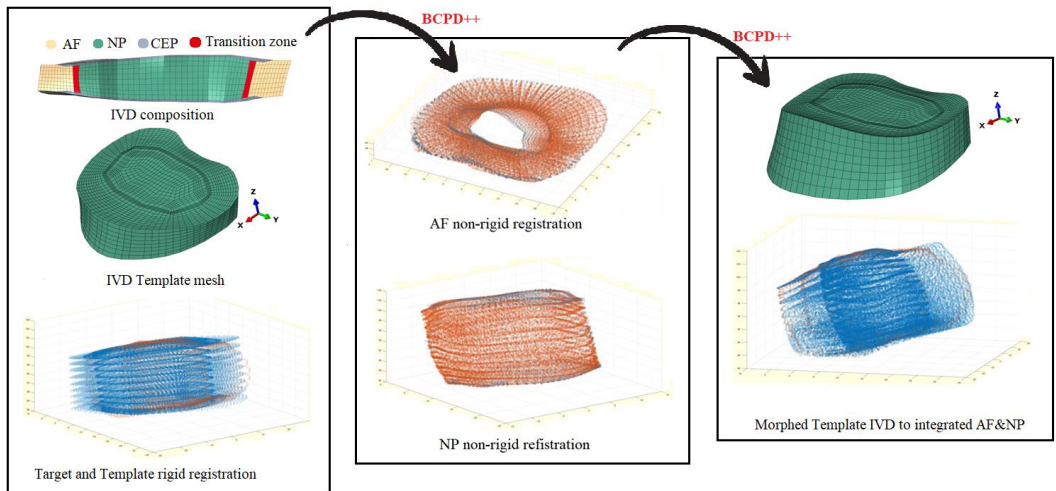

Figure 4: Stepwise IVD L5/L4 morphing process

*Thoracolumbar osteo-ligamentous hexahedral mean model*

34 thoracolumbar IVDs and vertebras, along with the pelvis, sacrum, and femoral head triangulated meshes, are integrated to create the complete model. Ligaments are modelled as elements by connecting corresponding points within the model. The ligaments are divided into six groups: Interspinous Ligament (ISL), Supraspinous Ligament (SSL), Ligamentum Flavum (LF), Capsular Ligament (CL), Intertransverse Ligament (ITL), and Posterior Longitudinal Ligament (PLL). The thoracolumbar osteo-ligamentous hexahedral mean model is depicted in Figure 5.

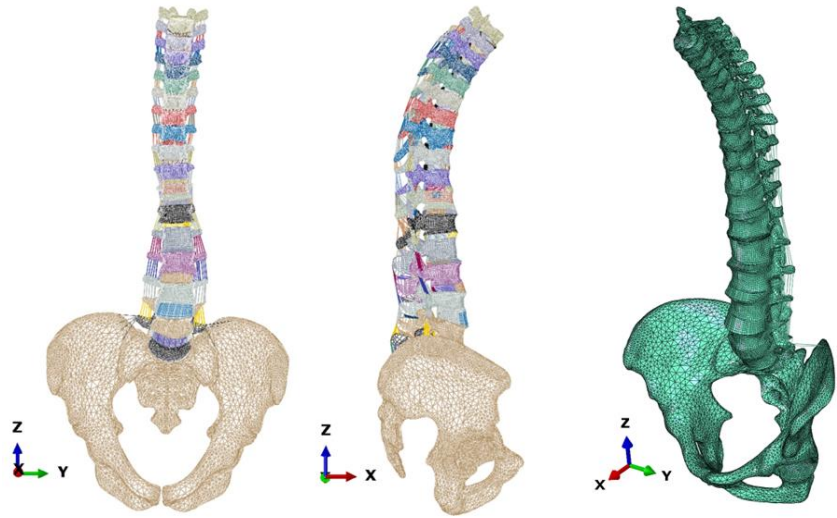

Figure 5: Thoracolumbar osteo-ligamentous hexahedral mean model

Soft tissue composition in thoracolumbar osteo-ligamentous hexahedral mean model is shown in Figure 6.

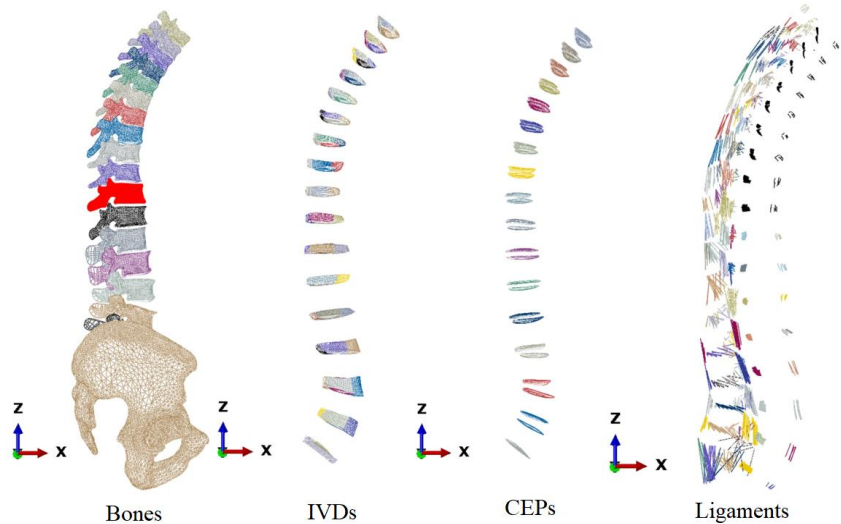

Figure 6: Soft tissue composition in thoracolumbar osteo-ligamentous hexahedral mean model

***Developing thoracolumbar osteo-ligamentous spine hexahedral SSM***

In the pipeline presented in Figure 1, the mean hexahedral model is aligned with the mean triangulated model using BCPD++ rigid registration with a large lambda value (1e9). The Iterative Closest Point (ICP) algorithm is then applied to find the closest corresponding points between the two mean models. By iteratively minimizing the distance between the points, the corresponding point IDs are obtained. Equation (4) is updated for the hexahedral SSM by replacing "Mean" with the mean hexahedral model and updating the prior matrix of shape deformations (sqrt(eigenvalues) * eigenvectors) for the corresponding points. This is achieved by considering the deformation vector of a point in the mean hexahedral mesh as the deformation vector of the closest point in the mean triangulated mesh (Figure 7a). Consequently, the updated equation (Shape = Hexahedral_Mean + DF × b) establishes a thoracolumbar spine hexahedral SSM tool (Figure 7b). In this updated equation, DF represents the updated deformation fields for the hexahedral meshes, and b corresponds to the standard deviations ranging from -3 to +3.

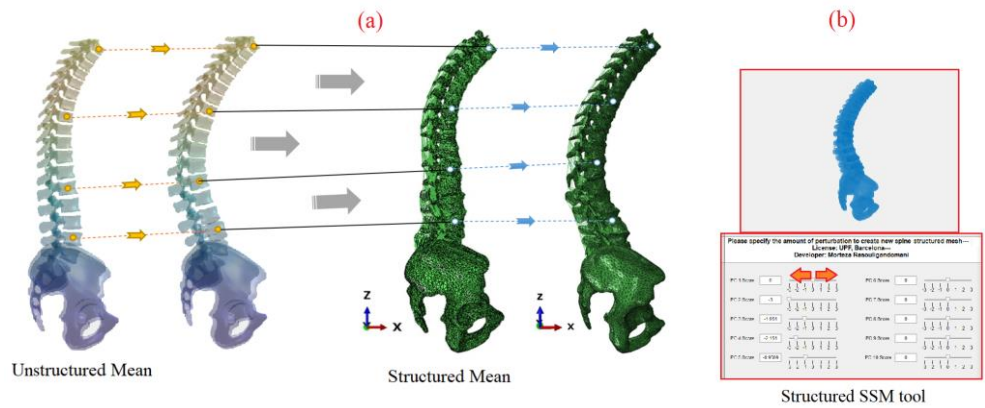

Figure 7: (a) Transferring triangulated mesh deformation vectors to hexahedral mesh correspondent points; (b) Automatized hexahedral SSM tool

16807 thoracolumbar spine osteo-ligamentous hexahedral FE models, comprising point coordinates, were sampled from the automated hexahedral SSM tool (Figure 7b). This was accomplished by combining the first five principal components (PCs) and adjusting the shape modes using seven standard deviations (-3, -2, -1, 0, 1, 2, 3) for each mode. The sampled data, along with the corresponding spinopelvic measurements, is stored in a public repository. Notably, the hexahedral SSM tool (Figure 7b) is independent of image data and relies solely on sagittal parameters of the spine, such as pelvic incidence (PI), pelvic tilt (PT), sacral slope (SS), and lumbar lordosis (LL). Geometrical parameters can be measured using clinical software like sterEOS or Surgimap. By activating different shape modes, various spine deformities can be obtained. A comparison between the hexahedral SSM and geometrical data through a pros and cons analysis can facilitate the utilization of patient-personalized FE thoracolumbar osteo-ligamentous spine models (Figure 8).

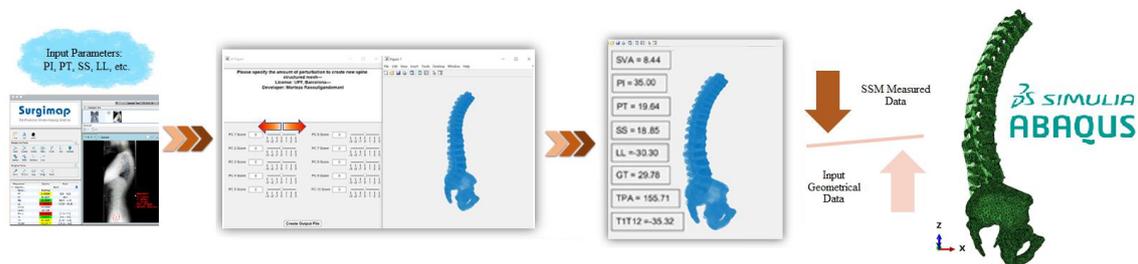

Figure 8: Pipeline to generate FE patient-personalized thoracolumbar osteo-ligamentous spine model using hexahedral SSM tool and input geometrical parameters

42 patient-personalized thoracolumbar spine hexahedral models were reconstructed based on the spinopelvic parameters of the included patients (pipeline in Figure 8). These models have been archived in a publicly accessible repository. Figure 9 showcases the first 10 examples of these patient-personalized thoracolumbar spine hexahedral models.

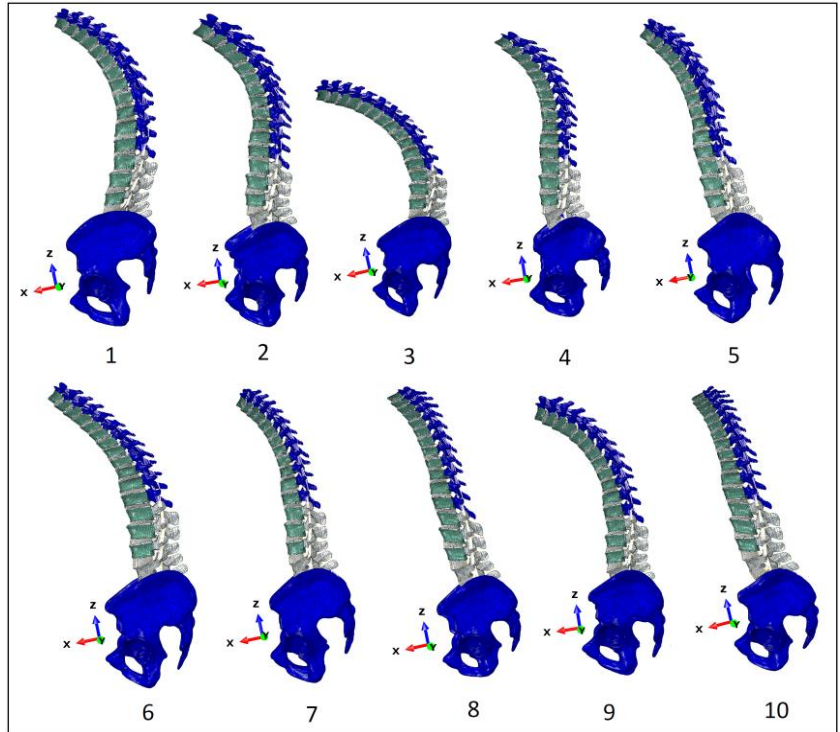

Figure 9: First 10 patient-personalized thoracolumbar spine hexahedral models out of 42 cases

## Data Records

The pipeline depicted in Figure 1 involves the sampling of 16807 triangulated and 16807 hexahedral thoracolumbar spine models utilizing the SSM and mesh morphing techniques. Furthermore, we provide 42 triangulated patient-personalized thoracolumbar spine EOS models obtained through sterEOS software, along with their corresponding 42 patient-personalized hexahedral models generated via the SSM. The DICOM images were not shared due to an ethical approval for collected bi-planar EOS images from IRCCS Milan, mentioning that the images could be used for scientific reasons, but sharing the DICOM files of patients was not allowed. All the virtual cohort models and 42 patient-personalized hexahedral models can be generated using 42 shared triangulated patient-personalized thoracolumbar spine EOS models as shared in the current repository. The spinopelvic geometrical parameters were evaluated for each model. All the data, 694GB and consisting of 12.7 billion nodes, are shared in a Zenodo repository as follows:

- *Thoracolumbar spine triangulated meshes[33]:*
- 42 stereolithography (stl) files (.stl extension) representing patient-personalized thoracolumbar spine triangulated meshes ("42 Patient-Specific stl files.rar"), including point coordinates and triangulated mesh connectivity IDs. Each stl file includes triangulated meshes of vertebras, pelvis, sacrum, and the femoral head[33].
- 16807 stl files representing the thoracolumbar spine triangulated models ("stl.part01.rar" to "stl.part09.rar"), including point coordinates and triangulated mesh connectivity IDs. Each stl file includes triangulated meshes of vertebras, pelvis, sacrum, and the femoral head[33].
- An excel file "Descriptive_List.xlsx" reporting measured spinopelvic parameters for 42 patient-personalized triangulated models, and virtual cohort 16807 triangulated models. Spinopelvic parameters are [14]: Pelvic-Incidence (PI), Pelvic Tilt (PT), Sacral Slope (SS), Lumbar Lordosis (LL), LL-PI, Global Tilt (GT), Relative Pelvic Version (RPV), Relative Lumbar Lordosis (RLL), Lumbar Distribution Index (LDI), Relative Spinopelvic Alignment (RSA), T1 pelvic Angle (TPA), and scoliosis cobb angle. Global Alignment and Proportion (GAP) score [34] is further measured as a complementary assessment. The Excel file has two sheets: sheet1 for 16807 virtual cohort, and sheet2 for 42 patient-personalized models[32]. Model ID in the excel file is correspondent to the model's name in both virtual cohort and patient-personalized models[33]. Model number in "Descriptive_List.xlsx" is correspondent to the same model number in 42 and 16807 FE input files, respectively (42 FE input files[35], 16807 virtual FE input files[36]).

- ***Thoracolumbar osteo-ligamentous spine hexahedral meshes[35,36]:***
- 42 FE input files (.inp extension, Abaqus software, Simulia) representing patient-personalized thoracolumbar spine hexahedral meshes, including point coordinates, mesh connectivity IDs and element sets. Each input file is almost 99MB (totally 3.86GB), and it includes vertebras and IVDs hexahedral meshes; pelvis, sacrum, and the femoral head triangulated meshes; and ligaments[35].
- One png file: "42_P-S.png" representing the first 10 patient-personalized thoracolumbar spine hexahedral models out of 42 FE models[35].
- 16807 FE input files representing thoracolumbar spine hexahedral models, including point coordinates. To reduce the size of shared virtual FE models, only point coordinates are shared here. The mean FE input file "Mean_Model (Template).inp" is also shared, which includes point coordinates, mesh connectivity IDs, and element sets. To generate virtual FE input files for any of the 16807 models, the corresponding shared point coordinates can be replaced into the mean FE input file ("Mean_Model (Template).inp"). Mesh connectivity IDs, and element sets are the same in all of FE input files. Mean FE input file includes vertebras and IVDs hexahedral meshes; pelvis, sacrum, and the femoral head triangulated meshes; and ligaments. Each point coordinate file is almost 39MB (totally 655GB)[36].
- Two excel files: "Descriptive_List (42_FE_virtual_models[35]; and 16807_FE_virtual_models[36]).xlsx" reporting measured spinopelvic parameters for 42 patient-personalized FE models, and 16807 virtual FE hexahedral models. The Excel file includes measured spinopelvic parameters (PI, PT, SS, LL, LL-PI, GT, RPV, RLL, LDI, RSA, TPA, and scoliosis cobb angle), GAP and IVD centric thickness for FE virtual cohort. Model ID in the excel file is correspondent to the model's name. Model number in " Descriptive_List (42_FE_virtual_models; and 16807_FE_virtual_models)" is correspondent to the same model number in 42 and 16807 stl files[33], respectively.
- One video file: "how_to_replace_point_coordinates.mp4". It shows how you can replace point coordinates here to the mean FE input file "Mean_Model (Template).inp" in order to generate specific FE input file[36].

## Technical Validation

The image datasets and thoracolumbar surface meshes used to train the triangulated SSM were obtained from IRCCS Galeazzi in Milan, utilizing the EOS modality and sterEOS software. Patient selection was based on predefined inclusion criteria. The 3D surfaces of the spines were reconstructed semi-automatically using the sterEOS software, with the assistance of an independent expert. The software enabled automatic validation of spine morphological parameters, and in case of any out-of-range values, the user had the ability to make necessary amendments to the reconstruction process. Additionally, morphological measurements were carefully reviewed and validated by a senior spine orthopedic surgeon at Hospital del Mar in Barcelona. This meticulous validation process ensured that the image-based reconstructed models achieved the highest level of accuracy and quality, as they underwent continuous measurements validation. Notably, the geometry re-calibration process was not required in this study, thanks to the utilization of EOS imaging technology.

It is important to emphasize that the validation of FE models is a crucial step in ensuring their accuracy and reliability. Range of Motion (ROM) analysis is a widely used technique for validating FE models, as it involves measuring the movement and displacement of a physical object.

In this study, the segmental ROM of the mean FE model was compared to previous works for validation purposes. The detailed description of constitutive laws falls beyond the scope of this paper. However, the material properties and boundary conditions employed for the ROM analyses are briefly outlined in Table 2.

Table 2: Material properties and boundary conditions for ROM analyses

| **Material properties** | *IVD* | *Vertebras* | *Ligaments* |
|---|---|---|---|
| | Anisotropic hyperelastic model [37] | Linear elastic [38] | Hypoelastic model [39] |
| **Boundary Conditions** | Sacrum is fixed, 7.5 N.m moment flexion, lateral bending, and axial rotation [40] | | |

The ROM of different spine segments was simulated using Abaqus software (Simulia) on a cluster supercomputer located at Pompeu Fabra University in Barcelona.

### *Segmental ROM: thoracic region*

The Functional Units (FU) of T10/T11, T6/T7, and T2/T3 were extracted, and pure moments of 7.5 N.m were applied to induce flexion-extension, lateral bending, and axial rotation in the upper endplates of each FU. The ROMs for the thoracic FUs are presented in Figure 10.

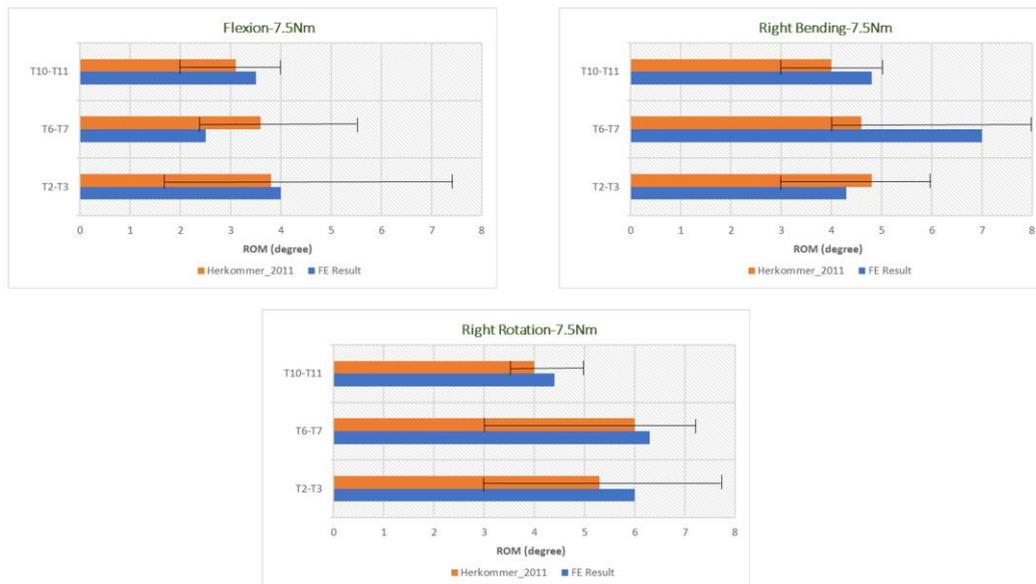

Figure 10: FU (T10/T11, T6/T7, T2/T3) ROM analyses; compared with [40]

The ROM obtained in this study are consistent with the findings of Herkommer et al. [40].

### *Segmental ROM: lumbar region*

The mean FE model from the pelvis to T10 is extracted, and pure moments of 7.5 N.m (flexion-extension, lateral bending, and axial rotation) are applied to the upper endplate of T10. The ROM in the lumbar region is evaluated for the segmental units L1/L2, L2/L3, L3/L4, and L4/L5. The lumbar FU ROMs are shown in Figure 11.

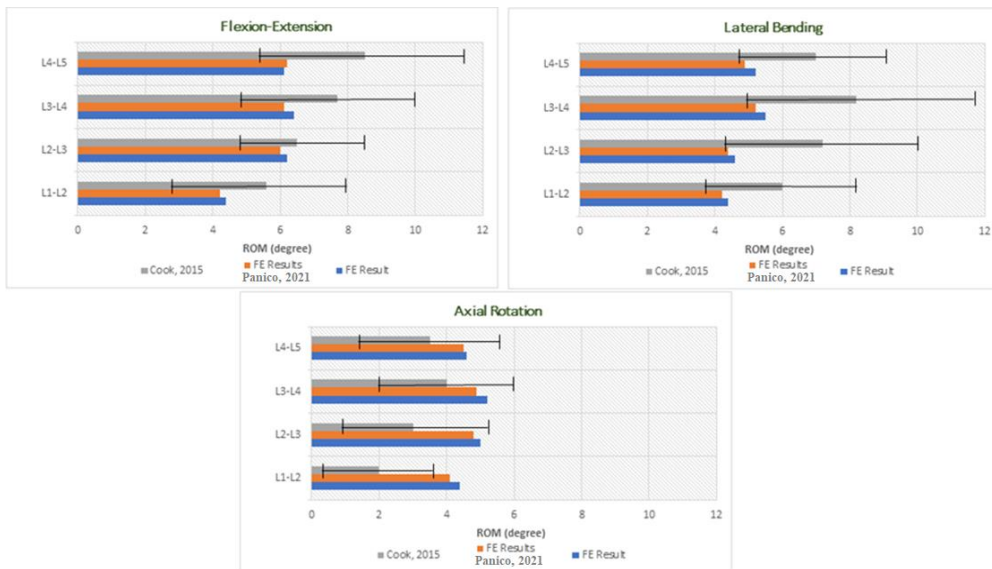

Figure 11: FU (L1/L2, L2/L3, L3/L4, L4/L5) ROM analyses; compared with [41, 42]

Results presented in Figure 11 are consistent with the findings of Cook [41] and Panico [42]. The validation of the thoracolumbar FE model is crucial for conducting subsequent biomechanical analyses and ASD treatment planning. Moreover, mesh quality of mean, 42, and 16807 FE models were assessed and validated in the following sections.

### Assessment of mean hexahedral mesh quality

The quality of the morphed mean hexahedral vertebras and IVDs, achieved through the BCPD++ method, was assessed and presented in Table 3. Among the mesh quality parameters, the Jacobian ratio was a commonly used measure in FE analyses, which evaluated the accuracy and reliability of numerical simulations [43-45]. A perfect finite element was defined in a reference system ($\xi_1$, $\xi_2$, $\xi_3$), where each point in the reference element was associated with its corresponding point in the modelled domain ($x_1$, $x_2$, $x_3$) through a mapping function F. The Jacobian matrix of the mapping function F, evaluated at a specific reference point $\xi$, was defined as $\frac{\partial F}{\partial \xi}(\xi)$ [46]. If the Jacobian was negative, the corresponding element could not be used for FE analysis. According to Oñate, Jacobian ratios between 0.3 and 0.8 were generally considered acceptable for most applications [47].

Table 3: Mean hexahedral mesh qualification for IVDs and vertebras

|      | *Mesh Error* | *Mesh Warnings* | *Jacobian ratio* | *Aspect Ratio > 10* | *Max Angle on Quad Faces > 160* |
|------|---|---|---|---|---|
| **Vertebras** | | | | | |
| *Min.* | 0% | 1.265% | 0.4 | 0.145% | 0.256% |
| *Max.* | 0% | 8.156% | 0.8 | 0.935% | 0.478% |
| **IVDs** | | | | | |
| *Min.* | 0% | 0.158% | 0.6 | 0.587% | 0.592% |
| *Max.* | 0% | 8.634% | 0.9 | 4.789% | 1.035% |
| **CEPs** | | | | | |
| *Min.* | 0% | 2.651% | 0.7 | 2.154% | 3.564% |
| *Max.* | 0% | 12.231% | 0.9 | 7,761% | 8.453% |

### Assessment and validation of mesh quality for virtual and patient-personalized FE models

To ensure that all elements in the deformed spine mesh (virtual and patient-personalized FE models) were valid for FE simulations, a post-processing step was performed on the mesh. Since the shape and elements of the spine might be distorted by the shape deformation applied through the SSM, it was important to evaluate the quality of the mesh. Specifically, the Jacobian of each hexahedral element was computed, and if it was found to be invalid (e.g., negative), further refinement was applied to validate the hexahedral mesh quality. For this purpose, the Sum-of-Squares (SOS) relaxation algorithm was employed to optimize the hexahedral meshes [48]. The available Matlab code from the GitHub repository[49] was used, and it required the MOSEK 9.3 Matlab optimization solver. In this study, 16807 virtual and 42 patient-personalized FE models were subjected for this mesh quality evaluation process. Remarkably, the percentage of invalid elements was 0%.

## Usage Notes

Clinicians and biomechanical researchers could utilize the provided excel file "Descriptive_List.xls" to search for a specific model of interest within the thoracolumbar spine morphology. Once identified, they could download the corresponding stl files and FE point coordinates. To generate FE input file for each of 16807 virtual models, the downloaded FE point coordinates could be substituted with the point coordinates in the mean FE input file "Mean_Model (Template).inp". A comprehensive usage guide for the shared data was presented in Figure 12.

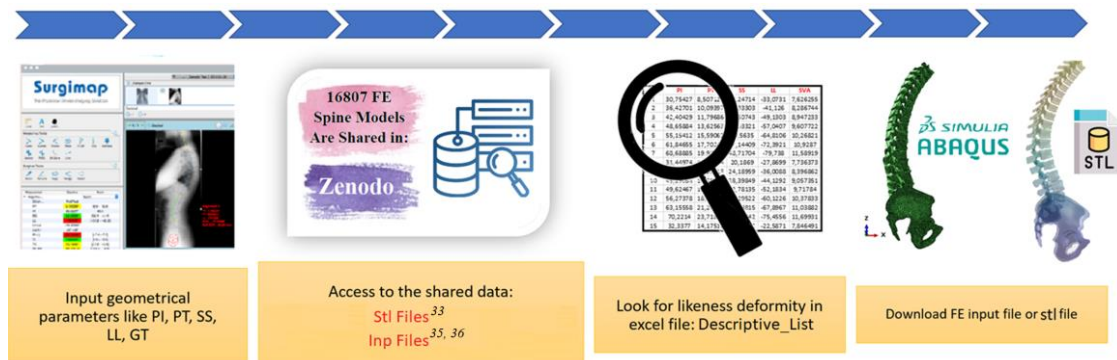

Figure 12: General usage guide for the shared data

The still files could be opened using various image viewers, 3D modelling, and CAD programs such as Microsoft 3D Viewer (for Windows) and MeshLab (available on multiple platforms). Besides, an online open-source platform under an A-GPL3 license has been developed to facilitate access to data, in which the data can be filtered, visualized, and downloaded through the data visualization platform[50].

Furthermore, the FE inp files could be opened using Abaqus 2019 and later versions. Other FE software that supported the inp extension could also open FE input files.

## Code Availability

The BCPD++ open-source codes could be accessed at the GitHub repository[32], and the SOS relaxation algorithm to check and repair the hexahedral meshes were available at the GitHub repository[49]. The repository for the online platform[50] can be accessed at [51].

## Acknowledgements

Funds received from DTIC-UPF, IMIM, IRCCS Istituto Ortopedico Galeazzi and Spanish Government (RYC-2015-18888, MDM-2015-0502).

## Author Contributions

MR wrote the manuscript, included 42 non-operated EOS bi-planar images, reconstructed the thoracolumbar spine 3D surfaces using the sterEOS software, developed the thoracolumbar spine triangulated and hexahedral SSM, sampled and shared the data in Zenodo repository. AdA revised the whole manuscript. FKC and MAB developed the online visualization platform. FG verified the thoracolumbar spine modelling using the sterEOS software (IRCCS, Milan, Italy). JN and MAGB revised the whole manuscript as well. All authors reviewed and confirmed the final manuscript.

## Competing interests
All authors declare no conflict of interest.